\documentclass[final,3p,times]{elsarticle}

\usepackage{amssymb}
\usepackage{amsmath}

\newcommand\mat\mathbf
\newcommand\dd{\, \mathrm{d}}
\newcommand\ii{\mathrm{i}}

\newcommand\um{~{\mu}\textrm{m}}
\newcommand\nm{\textrm{~nm}}
\newcommand\mm{\textrm{~mm}}

\makeatletter
\def\ps@pprintTitle{%
  \let\@oddhead\@empty
  \let\@evenhead\@empty
  \let\@oddfoot\@empty
  \let\@evenfoot\@oddfoot
}
\makeatother

\begin{document}

\begin{frontmatter}
	
\title{Design of cascaded diffractive optical elements generating different intensity distributions
at several operating wavelengths}

\author[label2,label1]{Georgy~A.~Motz}
\author[label2,label1]{Daniil~V.~Soshnikov}
\author[label1,label2]{Leonid~L.~Doskolovich}
\author[label2,label1]{Egor~V.~Byzov}
\author[label2,label1]{Evgeni~A.~Bezus}
\author[label1,label2]{Dmitry~A.~Bykov}

\affiliation[label2]{organization={Samara National Research University},
            addressline={34 Moskovskoye shosse}, 
            city={Samara},
            postcode={443086}, 
            country={Russia}}
						
\affiliation[label1]{organization={Image Processing Systems Institute, National Research Centre ``Kurchatov Institute''},
            addressline={151 Molodogvardeyskaya st.}, 
            city={Samara},
            postcode={443001}, 
            country={Russia}}

\begin{abstract}
We consider the design of cascaded diffractive optical elements (DOEs) for generating specified intensity distributions for several incident beams with different wavelengths.
For each incident beam with a given wavelength, the cascaded DOE forms a different specified intensity distribution.
The problem of DOE design is formulated as the problem of minimizing a functional that depends on the functions of diffractive microrelief height of the cascaded DOE and represents the deviations of the intensity distributions generated at the operating wavelengths from the specified ones.
Explicit expressions are obtained for the derivatives of the functional, and on this basis, a gradient method for calculating a cascaded DOE is formulated.
The proposed gradient method is used for the calculation of cascaded DOEs focusing radiation of three different wavelengths into different areas.
In particular, the calculation of a cascaded DOE generating a complex color image of parrots is considered. The presented numerical simulation results demonstrate high performance of the proposed method.
\end{abstract}

\begin{keyword}
diffractive optical element \sep inverse problem \sep scalar diffraction theory \sep gradient method
\end{keyword}

\end{frontmatter}

\section{Introduction}
Phase diffractive optical elements (DOEs) are used for solving various laser beam shaping and steering problems~\cite{1,2,3,4}.
Phase DOEs are usually implemented as a diffractive microrelief on a transmitting or reflecting substrate, which performs a certain phase modulation of the incident radiation. 
The calculation of a DOE is an inverse problem consisting of determining the function of the diffractive microrelief height (or of the phase modulation function proportional to the microrelief height), which ensures the formation of a light field with required parameters (e.\,g., with a required intensity distribution in a certain plane). 
Such calculations are usually performed in the framework of the scalar diffraction theory.
For designing DOEs, various iterative algorithms have been proposed, including the ``classical'' Gerchberg--Saxton algorithm, error-reduction algorithm, and a wide range of their modifications~\cite{5,6,7,8,9,10}.

It is important to note that in complex problems, including, for example, the generation of several different intensity distributions for several different incident beams (including those with different wavelengths), single DOEs demonstrate relatively low performance. 
To solve such problems, so-called cascaded DOEs are used, which consist of several sequentially placed DOEs and possess significantly wider functionality~\cite{2,11,12,13,14}.
In particular, in addition to solving problems of generating several different intensity distributions for different incident beams, cascaded DOEs have found wide application in various machine learning problems~\cite{14,15,16,17,18,19,20}, as well as in the problems of implementing various mathematical transformations described by linear operators~\cite{21,22,23}.
In these problems, due to a number of analogies between the cascaded DOEs and neural networks, the former are often referred to as the diffractive deep neural networks (DNNs)~\cite{15}.
The main method used for calculating the DNNs is the stochastic gradient descent method, as well as ``improved'' first-order methods based on it~\cite{24}.

Calculation of cascaded DOEs designed to work with radiation of several different wavelengths of great scientific and practical interest ~\cite{2,22,23,25,26,27,28}.
In what follows, we will refer to such cascaded DOEs (or DNNs) as the spectral ones. 
In particular, in recent works~\cite{22, 23}, spectral DNNs were considered for the optical implementation of various linear transformations at different wavelengths (each transformation being carried out at its own wavelength), as well as for the generation of multispectral images.
The calculation of spectral DNNs in~\cite{22, 23} was based on the stochastic gradient descent method, which demonstrated good performance. 
At the same time, gradient methods for the problem of calculating spectral DOEs that generate several different intensity distributions for incident beams with several different wavelengths remain insufficiently developed. For the sake of brevity, we will refer to this problem as the problem of focusing several different wavelengths (FSDW problem). 
The calculation of single and cascaded spectral DOEs for solving the FSDW problem was considered in~\cite{2, 11, 12, 24,25,26,27}.
In these works, iterative algorithms were used to calculate spectral DOEs, which are a generalization of the algorithms proposed for calculating “conventional” DOEs operating with radiation of a single wavelength. 
It is important to note that these algorithms are heuristic and do not have a strict theoretical basis. 
In the opinion of the present authors, the most significant results regarding the FSDW problem were obtained in~\cite{2}. 
In this paper, an iterative method was proposed for calculating cascaded spectral DOEs and applied for the calculation of cascaded DOEs generating different letters at different wavelengths. 
At the same time, the iterative calculation method proposed in~\cite{2} is also heuristic.
In particular, this method does not have the property of not increasing the error at each iteration, which the Gerchberg--Saxton algorithm and the error-reduction algorithm possess. 
Thus, the development and investigation of a gradient method for calculating spectral DOEs for the FSDW problem is of great interest.

In the present work, for solving the FSDW problem, we consider a gradient method for calculating a cascaded spectral DOE. 
In the method, the problem of calculating a DOE is formulated as the problem of minimizing an error functional that depends on the functions of the diffractive microrelief height of the cascaded DOE and describes the deviation of the intensity distributions generated at the design wavelengths from the required ones. 
Explicit and compact expressions are obtained for the derivatives of the error functional with respect to the microrelief height functions. 
Using the proposed gradient method, several examples of single and cascaded spectral DOEs are designed and numerically investigated.

\section{Statement of the problem of calculating a cascaded spectral DOE}\label{sec:1}

\begin{figure}[hbt]
	\centering
		\includegraphics{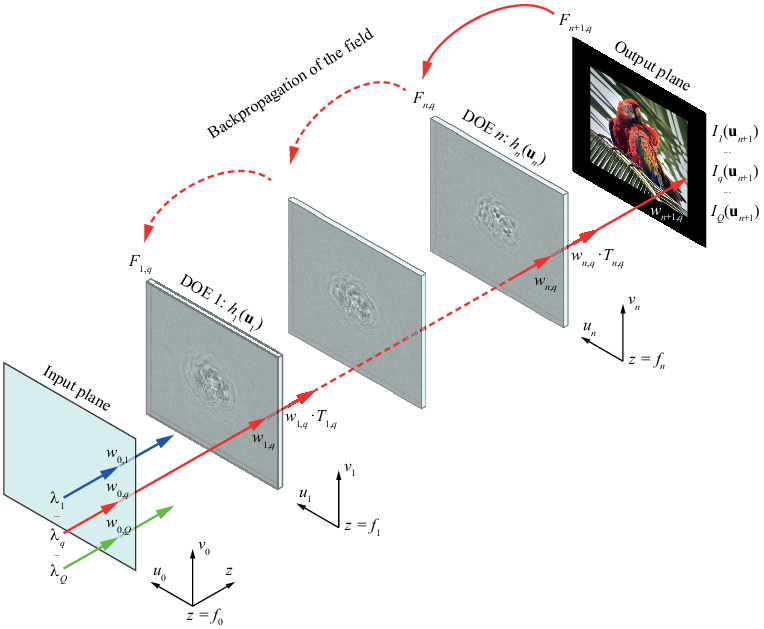}
	\caption{\label{fig:1}Geometry of the problem of calculating a cascaded spectral DOE.}
\end{figure}

Let us consider the problem of calculating a cascaded spectral DOE generating prescribed intensity distributions for $Q$ different incident beams having different wavelengths ${\lambda_q},\; q = 1,\ldots,Q$.
We assume that the cascaded DOE consists of $n$ phase DOEs located in the planes $z = f_1, \ldots, z = f_n$ $(0 < f_1 < \cdots < f_n)$ and defined by the functions of diffractive microrelief height $h_1(\mat{u}_1), \ldots , h_n( \mat{u}_n )$, where $\mat{u}_j = ( u_j, v_j )$ are Cartesian coordinates in the plane $z = f_j$ (Fig.~\ref{fig:1}). 

Let for each wavelength $\lambda_q \in \{ \lambda_1, \ldots, \lambda_Q \}$, an input field (incident beam) with complex amplitude $w_{0,q}(\mat{u}_0)$ be defined in the input plane $z = f_0 = 0$.
Here and in what follows, the subscripts of the complex amplitude of the field denote the index of the plane, in which this amplitude is written, and the index of the wavelength. 
We assume that the free-space propagation of the light field between the DOEs is described by the Fresnel--Kirchhoff diffraction integral, and that the transmission of the field through a DOE is described by the multiplication of its complex amplitude by the complex transmission function (CTF) of the DOE. 
The CTF of the $m$-th DOE is wavelength-dependent and, at a certain wavelength $\lambda_q$, reads as
\begin{equation}
\label{eq:1}
T_{m,q}(\mat{u}_m) 
= \exp \left\{ \ii \varphi_{m,q}(\mat{u}_m) \right\} 
= \exp \left\{ \ii \frac{2\pi}{\lambda_q} [n(\lambda_q) - 1] h_m( \mat{u}_m ) \right\},
\end{equation}
where $\varphi_{m,q}(\mat{u}_m)$ is the phase function of the $m$-th DOE (the phase shift introduced by it) at the wavelength $\lambda_q$ and $n(\lambda_q)$ is the refractive index of the DOE material.
Under the made assumptions, the propagation of the incident beams $w_{0,q}( \mat{u}_0 )$, $q = 1,\ldots ,Q$ from the input plane $z = 0$ through the cascaded DOE to the output plane $z = f_{n+1}$ is described by the following recurrent formula:
\begin{equation}
\label{eq:2}
\begin{aligned}
w_{m,q}( \mat{u}_m ) = C_{m,q}\iint w_{m-1, q}(\mat{u}_{m-1}) T_{m-1,q}( \mat{u}_{m-1})
\cdot
&\exp \left\{ \ii\frac{\pi}{\lambda_q d_m}{( \mat{u}_m  - \mat{u}_{m-1} )^2} \right\} \dd^2 \mat{u}_{m-1} ,
\\&\hspace{9em}m = 1,\ldots ,n+1,
\end{aligned}
\end{equation}
where $w_{m,q}( \mat{u}_m ),\,m = 1,\ldots ,n$ is the complex amplitude of the field incident on the $m$-th DOE with the CTF $T_{m,q}(\mat{u}_m )$, $C_{m,q} = (\ii\lambda _q d_m )^{-1} \exp \{ \ii 2\pi d_m/\lambda_q \}$, and $d_m = f_m - f_{m-1}$ is the distance between the $m$-th DOE and the previous one.
Let us note that for the calculation of the fields $w_{1,q}( \mat{u}_1 )$ incident on the first DOE [$m=1$ in Eq.~\eqref{eq:2}], one should use $T_{0,q}(\mat{u}_0) = 1$.
Here, the integrals are taken over the aperture of the corresponding DOE (at $m>1$) or over the input fields (at $m = 1$).

Let us now consider the inverse problem of calculating a cascaded DOE. Under this problem, we will understand the problem of calculating the functions of the diffractive microrelief height $h_1( \mat{u}_1 ), \ldots, h_n( \mat{u}_n )$ of the cascaded DOE, which, for the given input beams $w_{0,q}( \mat{u}_0 ),\,\,q = 1,\ldots ,Q$, provide the generation of light fields with the required intensity distributions $I_q(\mat{u}_{n+1} ),\,\,q = 1,\ldots ,Q$ in the output plane. 
We will assume that the error of the generation of a required intensity distribution $I_q( \mat{u}_{n+1})$ for the corresponding incident beam $w_{0,q}( \mat{u}_0 )$ with the wavelength $\lambda_q$ is represented by a certain integral criterion
\begin{equation}
\label{eq:3}
\varepsilon_q ( h_1, \ldots, h_n ) = \iint D_q\left( I_{n+1,q} ( \mat{u}_{n+1} ), I_q(\mat{u}_{n+1}) \right) \dd^2 \mat{u}_{n+1},
\end{equation}
where $I_{n+1,q}( \mat{u}_{n+1} ) = \left| w_{n+1,q}( \mat{u}_{n+1} ) \right|^2$ is the intensity distribution generated in the output plane at the functions $h_1( \mat{u}_1 ), \ldots, h_n( \mat{u}_n  )$, and $D_q$ is a certain function describing the difference between the generated and required distributions at the current point. 
Then, the inverse problem of calculating a cascaded DOE generating required intensity distributions $I_q(\mat{u}_{n+1} ),\,\,q = 1, \ldots, Q$ for all incident beams $w_{0,q}( \mat{u}_0 ), \,\,q = 1,\ldots, Q$ with different wavelengths can be considered as a problem of minimization of the sum of the presented functionals
\begin{equation}
\label{eq:4}
\varepsilon ( h_1, \ldots, h_n ) = \sum_{q = 1}^Q \varepsilon_q( h_1, \ldots, h_n )  \stackrel{h_1, \ldots, h_n}{\longrightarrow} \min. 
\end{equation}

\section{Gradient method for calculating a cascaded spectral DOE}
For the functional of Eq.~\eqref{eq:4}, it is easy to calculate the Fréchet derivatives 
$\delta \varepsilon ( h_1, \ldots, h_n ) / \delta h_m$, which makes it possible to use a gradient method for solving the problem of Eq.~\eqref{eq:4}. 
Indeed, since the functional of Eq.~\eqref{eq:4} equals the sum of functionals, its derivatives have the form
\begin{equation}
\label{eq:5}
\frac{\delta \varepsilon (h_1, \ldots, h_n )}{\delta h_m} 
= \sum\limits_{q=1}^Q {\frac{\delta \varepsilon_q( h_1, \ldots, h_n )}{\delta h_m}} ,\; m = 1, \ldots, n.
\end{equation}

The calculation of a derivative of a functional $\delta \varepsilon_q( h_1, \ldots, h_n ) / \delta h_m$ in~\eqref{eq:5} is carried out similarly to the problem of calculating a cascaded DOE generating a required intensity distribution in the case of a single incident beam $w_{0,q}( \mat{u}_0 )$ with the wavelength $\lambda_q$. 
A detailed description of the calculation of such derivatives is given in a recent work~\cite{14} by the present authors. 
According to~\cite{14}, this derivative can be easily obtained using the unitarity property of the operator describing the light propagation through a cascaded DOE in the form
\begin{equation}
\label{eq:6}
\frac{\delta \varepsilon_q( h_1, \ldots, h_n )}{\delta h_m} 
= -2\gamma_q\operatorname{Im} \left[w_{m,q}(\mat{u}_m) T_{m,q}(\mat{u}_m)  F_{m,q}^*(\mat{u}_m)\right],
\end{equation}
where $\gamma_q = 2\pi [ n(\lambda_q) - 1 ] / \lambda_q$, $w_{m,q}( \mat{u}_m )$ is the complex amplitude of the field incident on the $m$-th DOE upon the forward propagation of the field $w_{0,q}( \mat{u}_0 )$ from the plane $z=0$ to the plane $z = f_m$ [the calculation of the field $w_{m,q}( \mat{u}_m )$ is carried out using Eq.~\eqref{eq:2}], $T_{m,q}(\mat{u}_m )$ is the CTF of the $m$-th DOE at the wavelength ${\lambda_q}$ defined by Eq.~\eqref{eq:1}, and $F_{m,q} ( \mat{u}_m )$ is the complex amplitude of the field obtained upon the ``backpropagation'' of the so-called error field $F_{n+1,q}( \mat{u}_{n+1} )$
from the output plane $z = f_{n+1}$ to the plane $z = {f_m}$.
The error field has the following form~\cite{14}:
\begin{equation}
\label{eq:7}
F_{n+1,q}( \mat{u}_{n+1} ) = \frac{\partial {D_q}\left(I_{n+1,q}( \mat{u}_{n+1} ), I_q(\mat{u}_{n+1}) \right)}{\partial I_{n+1,q}}
w_{n+1,q} ( \mat{u}_{n+1} ).
\end{equation}

Note that the backpropagation of the field in the free space is also described by the Fresnel--Kirchhoff integral, in which the propagation distance is taken with a minus sign, and the ``reverse transmission'' of the beam through a DOE is described by the multiplication of the complex amplitude of the beam by the complex conjugate of the CTF of this DOE. 
Thus, the function $F_{m,q} ( \mat{u}_m )$ in Eq.~\eqref{eq:6} is calculated using the following formula:
\begin{equation}
\label{eq:9}
\begin{aligned}
F_{l,q}( \mat{u}_{l} ) = 
C^*_{l+1,q} \iint F_{l+1,q}(\mat{u}_{l+1})  T_{l+1,q}^*(\mat{u}_{l+1})
\exp &\left\{ -\ii\frac{\pi}{\lambda_q d_{l+1}} ( \mat{u}_{l} - \mat{u}_{l+1} )^2 \right\}
\dd^2 \mat{u}_{l+1}.
\end{aligned}
\end{equation}
This equation is used recursively starting from $l = n$ down to $l = m$. 
To use Eq.~\eqref{eq:9} at $l = n$, we have to take $T_{n+1,q}^*(\mat{u}_{n+1}) = 1$.

The presented expressions~\eqref{eq:5}--\eqref{eq:7} for calculating the derivatives of the error functional constitute the basis for solving the inverse problem of Eq.~\eqref{eq:4} with a gradient method or a certain improved first-order method, e.g., ADAM~\cite{24}. 
Let us note that these expressions were obtained for error functional $\varepsilon_q ( h_1, \ldots, h_n)$ written in a general form~\eqref{eq:3} and depending on the function $D_q$. 
In an important particular case of $D_q\left( I_{n+1,q}( \mat{u}_{n+1} ), I_q ( \mat{u}_{n+1} ) \right) = \left[ I_{n+1,q} (\mat{u}_{n+1} ) - {I_q}(\mat{u}_{n+1} ) \right]^2$, the functionals~\eqref{eq:3} turn into integrals of the squared error:
\begin{equation}
\label{eq:10}
\varepsilon_q ( h_1, \ldots, h_n ) = \iint \left[ {I_{n+1,q}(\mat{u}_{n+1} ) - I_q(\mat{u}_{n+1} )} \right]^2 \dd^2\mat{u}_{n+1} .
\end{equation}
In this case, the derivatives $\delta\varepsilon_q( h_1, \ldots, h_n ) / \delta h_m$ are calculated using the general formula~\eqref{eq:6}, and the error field has the form
\begin{equation}
\label{eq:11}
F_{n+1,q}( \mat{u}_{n+1} ) = 2\left[ I_{n+1,q}(\mat{u}_{n+1} ) - I_q(\mat{u}_{n+1} ) \right] w_{n+1,q}(\mat{u}_{n+1} ).
\end{equation}

Let us note that the functions of the diffractive microrelief height $h_1( \mat{u}_1 ), \ldots, h_n( \mat{u}_n )$ are usually supposed to be defined in a certain interval $[0, h_{\rm max})$, where $h_{\rm max}$ is the maximum height of the diffractive microrelief (the $h_{\rm max}$ value is chosen taking into account the technology used for the DOE fabrication). 
The presence of the constraints $0 \leqslant h_m( \mat{u}_1 ) \leqslant h_{\rm max},\,\,i = 1,\ldots ,n$ makes the problem of calculating a cascaded DOE a conditional optimization problem. 
In order to take into account the presented constraints, the following projection operator on the set of ``limited height'' functions has to be introduced to the iterative process of calculating the functions of the diffractive microrelief heights:
\begin{equation}
\label{eq:12}
{\rm P} ( h ) = 
\begin{cases}
  0, &h < 0, \\ 
  h, &h \in [0, h_{\rm max}), \\ 
  h_{\rm max}, &h \geqslant h_{\rm max}. 
\end{cases}
\end{equation}

In particular, the introduction of the operator~\eqref{eq:12} to the gradient method for calculating a cascaded DOE corresponds to the gradient projection method, in which the calculation of the next set of the microrelief height functions is carried out using the expression
\begin{equation}
\label{eq:13}
h_m^k( \mat{u}_m ) = {\rm P} \left( h_m^{k-1}( \mat{u}_m) - t\frac{\delta \varepsilon}{\delta h_m}( \mat{u}_m ) \right),\,\,m = 1,\ldots ,n,
\end{equation}
where the superscript $k$ denotes the iteration number and $t$ is the step of the gradient method.

\section{Design examples of cascaded spectral DOEs}
Since the cascaded spectral DOEs are calculated from the condition of generating a different prescribed intensity distribution $I_q( \mat{u}_{n+1} )$ at each working wavelength $\lambda_q$, then, in the case of an incident beam $\sum_q w_{0,q}( \mat{u}_0 )$ consisting of input beams with the operating wavelengths, the cascaded DOE will generate a color image corresponding to the sum of monochromatic intensity distributions $I_q( \mat{u}_{n+1} ),\,q = 1, \ldots, Q$. In this regard, below we consider two design examples of DOEs generating such color images.

\begin{figure}
	\hspace{-4em}
		\includegraphics{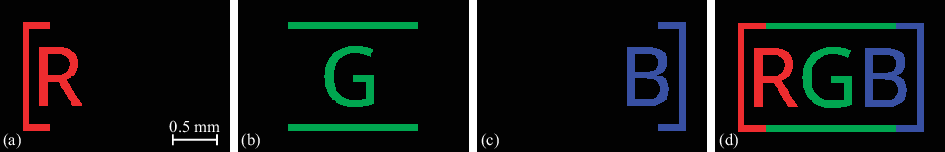}
	\caption{\label{fig:2}Required intensity distributions for the three design wavelengths (a)--(c) and the resulting color ``RGB-distribution'' (d).}
\end{figure}

\paragraph{Example 1}
Let us design a cascaded DOE operating at three wavelengths $\lambda_1 = 633\nm$, $\lambda_2 = 532\nm$, and $\lambda_3 = 457\nm$ from the red, green, and blue parts of the visible spectrum and generating a color image consisting of the letters ``R'', ``G'', and ``B'' in a frame (Fig.~\ref{fig:2}).
The required intensity distributions $I_q( \mat{u}_{n+1} ),\,\,q = 1,2,3$, which have to be generated by the cascaded DOE at the chosen wavelengths, are shown in Figs.~\ref{fig:2}(a)--\ref{fig:2}(c) with different colors and correspond to the images of different letters and fragments of the frame.
In the design, we assume that the input fields are Gaussian beams $w_{0,q}( \mat{u}_0 ) = A_q\exp \{ -\mat{u}_0^2/{\sigma ^2}\},\,\,q = 1,2,3$ with $\sigma = 1.2\mm$ and the amplitudes $A_q$ chosen from the following normalization condition:
\begin{equation}
\label{eq:14}
\iint \left| w_{0,q}( \mat{u}_0 ) \right|^2  \dd^2\mat{u}_0
= \iint I_q(\mat{u}_{n+1} )\dd^2 \mat{u}_{n+1},
\end{equation}
which provides the equality of the energy of the input beam with the wavelength $\lambda_q$ and the corresponding required intensity distribution $I_q( \mat{u}_{n+1} )$.

The calculation of the microrelief height functions of the cascaded DOEs generating the required color RGB-distribution was carried out using the gradient method of Eqs.~\eqref{eq:5}--\eqref{eq:13} described above.
As the functionals $\varepsilon_q( h_1, \ldots, h_n )$ describing the error of the generation of the required intensity distributions at the design wavelengths, the ``squared error functionals'' of Eq.~\eqref{eq:10} were used.
In this case, the derivatives of the functionals $\varepsilon_q( h_1, \ldots, h_n )$ are calculated using the general Eq.~\eqref{eq:6}, where the function $F_{m,q}( \mat{u}_m )$ is obtained through the backpropagation of the error field~\eqref{eq:11}. 
Let us note that in the present work, the fields used in Eq.~\eqref{eq:6} were calculated using the angular spectrum method~\cite{29, 30}.

In the present example, we designed a single DOE and cascaded DOEs consisting of two and three DOEs.
In the calculations, it was assumed that the distance from the input plane to the first DOE, the distances between DOEs, and the distance from the last DOE to the output plane are the same and equal 80~mm.
The microrelief height functions in the DOE planes were defined on $512 \times 512$ square grids with a $10\um$ step.
In this case, the aperture of each DOE was a square with the sides equal to 5.12~mm.
The fields in the input and output planes were defined on $1024 \times 1024$ grids with the same step.
The maximum height of the diffractive microrelief was taken to be equal to $h_{\rm max} = 6\um$. 
Note that DOEs with such height can be fabricated using the standard direct laser writing technique~\cite{10, 31}.
As the refractive indices of the DOE material, the values $n(\lambda_1) = 1.457$, $n(\lambda_2) = 1.461$, and $n(\lambda_3) = 1.465$ corresponding to fused silica were chosen.

\begin{figure}
	\hspace{-3em}
		\includegraphics{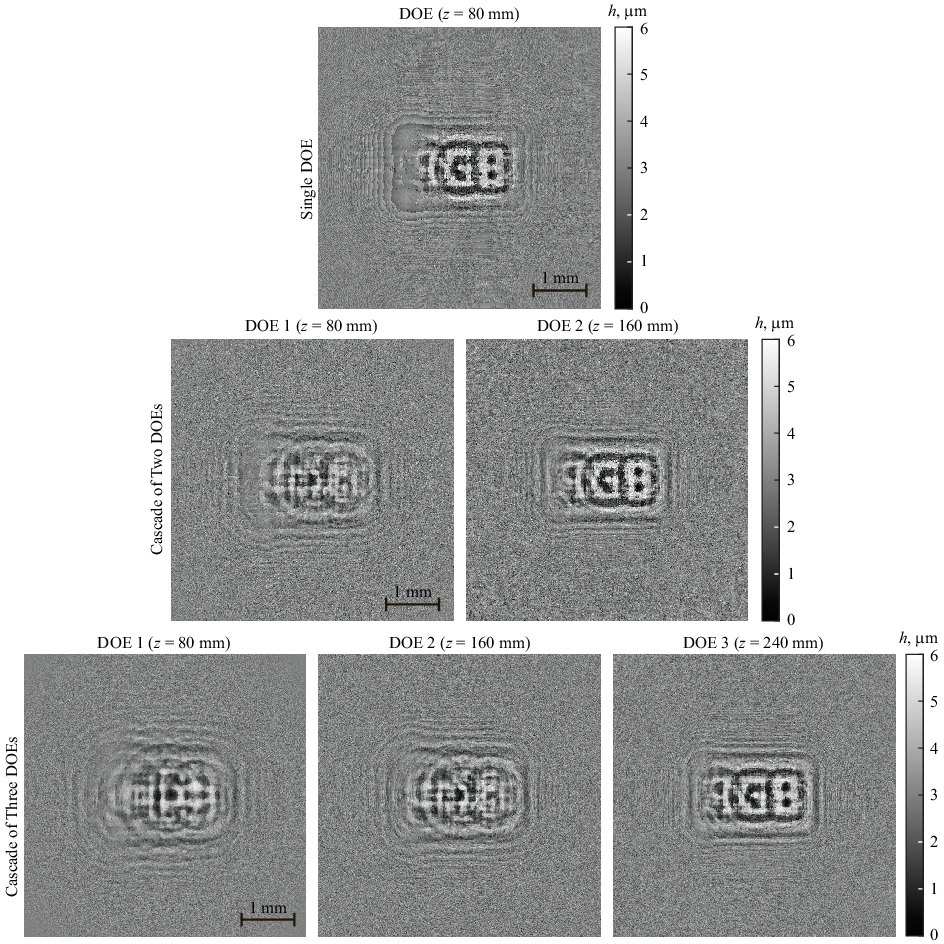}
	\caption{\label{fig:3}Calculated microrelief height functions of the designed single DOE (top row), cascade of two DOEs (middle row), and cascade of three DOEs (bottom row).}
\end{figure}

The calculated microrelief height functions of the designed single and cascaded DOEs are shown in Fig.~\ref{fig:3}.
In the calculations, 5000~iterations were made in each case, with an exponentially decreasing step (such a number of iterations turned out to be sufficient for the convergence of the method). 
As the initial microrelief height functions, realizations of a white noise signal with a uniform distribution in the range $[0, h_{\rm max})$ were used. 
The computation time on an NVIDIA RTX 3060 12~Gb graphics card was from 4--5 minutes for a single DOE to approximately 15~minutes for the cascade of three DOEs.

\begin{figure}
	\hspace{-3em}
		\includegraphics{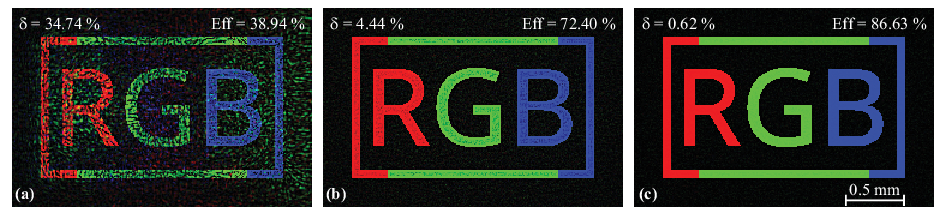}
	\caption{\label{fig:4}Calculated color ``RGB-distributions'' generated by the designed single DOE (a), cascade of two DOEs (b), and cascade of three DOEs (c). Over the distributions, the mean energy efficiencies and root-mean-square deviations at the design wavelengths are shown.}
\end{figure}

Figure~\ref{fig:4} shows the intensity distributions generated by the calculated single and cascaded DOEs.
These distributions
correspond to the sum of the intensity distributions $I_{n+1,q}(\mat{u}_{n+1})$ generated by the designed DOEs at the operating wavelengths $\lambda_q$. 
From Fig.~\ref{fig:4}, it is evident that the quality of the generated distributions rapidly increases with an increase in the number of elements constituting the cascaded DOE.
In particular, for a single DOE [Fig.~\ref{fig:4}(a)], the quality of the generated color distribution is rather poor and contains pronounced noise components.
In contrast, the distribution of Fig.~\ref{fig:4}(c) generated by a cascade of three DOEs is already visually indistinguishable from the required distribution shown in Fig.~\ref{fig:2}(d).

In order to characterize the quality of the generated distributions in a quantitative form, let us introduce the energy efficiency ${\rm Eff}$ and the root-mean-square deviation $\delta$.
Let us define the energy efficiency as
\begin{equation}
\label{eq:15}
{\rm Eff} = \frac1Q \sum\limits_q {\rm Eff}_q  = \frac1Q \sum\limits_q \frac1{E_{0,q}}  \iint_{G_q} I_{n+1,q}(\mat{u}_{n+1})\dd^2\mat{u}_{n+1},
\end{equation}
where ${\rm Eff}_q$ corresponds to the fraction of energy $E_{0,q} = \iint | w_{0,q}(\mat{u}_0) |^2 \dd^2\mat{u}_0$ of the incident beam with the wavelength $\lambda_q$, which reaches the region corresponding to the required distribution $G_q = \left\{ {\left. \mat{u}_{n+1}  \right|\,\,I_q(\mat{u}_{n+1} ) \ne 0} \right\}$ for this wavelength.
Thus, the energy efficiency ${\rm Eff}$ describes the average energy fraction of the incident beams, which is directed to the required regions. Similarly, the quantity
\begin{equation}
\label{eq:16}
\delta 
= \frac1Q \sum\limits_q \delta_q 
= \frac1Q \sum\limits_q \frac1{M_q} \sqrt {\iint_{G_q} \left[I_{n+1,q}(\mat{u}_{n+1}) - {\rm Eff}_q \cdot {I_q}(\mat{u}_{n+1})\right]^2\dd^2{\mat{u}_{n+1}}} 
\end{equation}
describes the averaged root-mean-square deviation of the intensity distributions $I_{n+1,q} (\mat{u}_{n+1})$ generated at the operating wavelengths from the required distributions $I_q(\mat{u}_{n+1})$. 
Note that the root-mean-square deviations ${\delta_q}$ for the wavelengths ${\lambda_q}$ are normalized by the average values 
$$
M_q = \| G_q \|^{-1}  \iint_{G_q} I_{n+1,q}(\mat{u}_{n+1} )  \dd^2\mat{u}_{n+1} ,
$$
where $\| G_q \|$ is the area of the region $G_q$. 
The obtained values of the energy efficiency and root-mean-square deviation in percent for the designed DOE examples are shown in Fig.~\ref{fig:4} over each of the generated distributions. 
From the presented values, one can see that for the cascade of three DOEs, the root-mean-square error $\delta $ does not exceed 0.7\%, and the energy efficiency is over 86\%.

\begin{figure}
	\hspace{-3em}
		\includegraphics{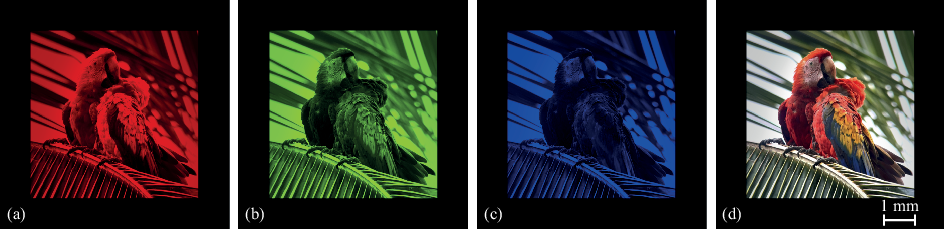}
	\caption{\label{fig:5}Required intensity distributions for the three design wavelengths~(a)--(c) and the resulting color image of parrots~(d).}
\end{figure}

\paragraph{Example 2} Let us now consider a more complex problem of calculating a cascaded DOE generating a color image of parrots. 
We assume that the required color image is a sum of three monochromatic intensity distributions shown in Figs.~\ref{fig:5}(a)--\ref{fig:5}(c) with the wavelengths $\lambda_1 = 633\nm$, $\lambda_2 = 532\nm$, and $\lambda_3 = 457\nm$ (these monochromatic distributions are defined on $512 \times 512$ grids with the step of $10\um$). 
In this case, the resulting color distribution
will correspond to the image shown in Fig.~\ref{fig:5}(d)~\cite{32}. 
Accordingly, the problem of calculating a cascaded DOE for generating the color image of Fig.~\ref{fig:5}(d) can be considered as a problem of generating intensity distributions of Figs.~\ref{fig:5}(a)--\ref{fig:5}(c) at the three presented operating wavelengths.

\begin{figure}
	\hspace{-3em}
		\includegraphics{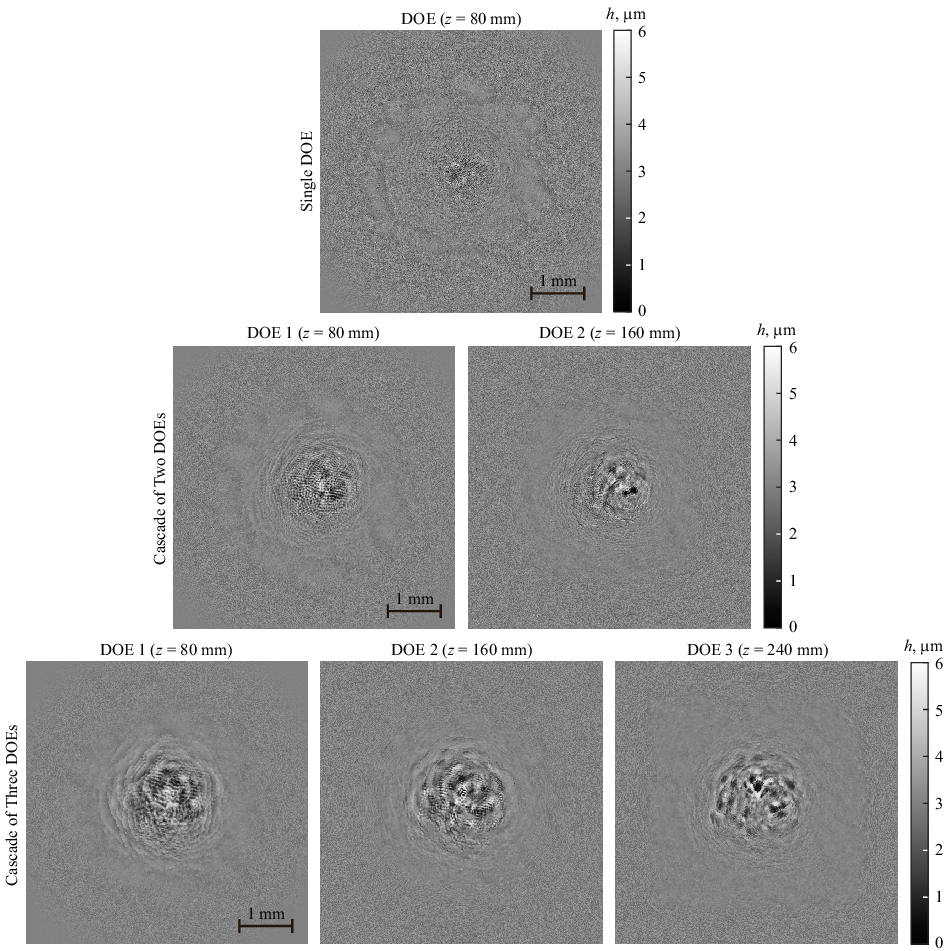}
	\caption{\label{fig:6}Calculated microrelief height functions of the designed single DOE (top row), cascade of two DOEs (middle row), and cascade of three DOEs (bottom row) generating a color image of parrots.}
\end{figure}

\begin{figure}
	\hspace{-3em}
		\includegraphics{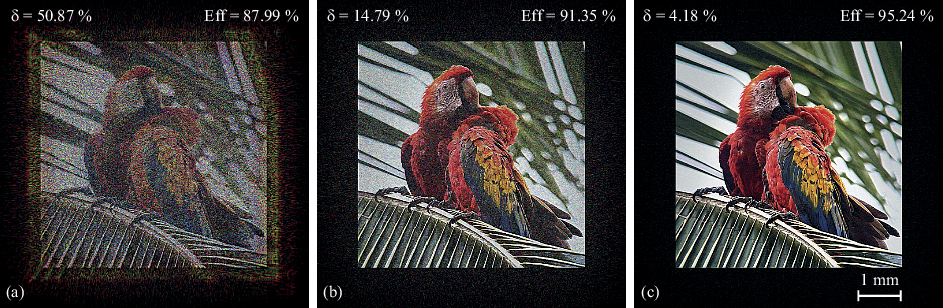}
	\caption{\label{fig:7}Calculated color images of parrots generated by the designed single DOE (a), cascade of two DOEs (b), and cascade of three DOEs (c). Over the distributions, the mean energy efficiencies and root-mean-square deviations are shown.}
\end{figure}

The calculation of the microrelief height functions of the cascaded DOEs generating the color image of parrots was carried out using the proposed gradient method.
The incident beams $w_{0,q}( \mat{u}_0 )$, $q = 1,2,3$ as well as all the other parameters (number of DOEs, distances between planes, discretization and size of DOEs, maximum microrelief height, and refractive indices of the DOE material) were the same as in example~1 considered above.
The calculated microrelief height functions of the designed single DOE and cascades of two and three DOEs are shown in Fig.~\ref{fig:6}.
Figure~\ref{fig:7} shows the intensity distributions generated by the calculated DOEs.
As in the previous example, these distributions 
correspond to the sum of distributions $I_{n+1,q}(\mat{u}_{n+1} )$ generated by the designed DOEs at the operating wavelengths $\lambda_q$.
Figure~\ref{fig:7} shows that the quality of the generated color distribution becomes reasonably good already in the case of the cascade of two DOEs.
In this case, the energy efficiency exceeds 91\%, whereas the root-mean-square deviation is less than 17\%.
For the cascade of three DOEs, the generated distribution [Fig.~\ref{fig:7}(c)] becomes visually indistinguishable from the required color image shown in Fig.~\ref{fig:5}(d), the energy efficiency is greater than 95\%, and the root-mean-square deviation does not exceed 4.2\%, which is unachievable with a single DOE [see Fig.~\ref{fig:7}(a)].

\section{Conclusion}
In this work, we presented a gradient method for calculating cascaded DOEs generating several required intensity distributions for several incident beams with different wavelengths. 
The problem of calculating a cascaded DOE was formulated as a problem of minimizing a functional depending on the functions of the diffractive microrelief height of the cascaded DOE and representing the error of the generation of the required distributions at the design wavelengths. 
Explicit and compact expressions were obtained for the derivatives of this functional. 
Using the proposed gradient method, we calculated single and cascaded DOEs generating specified intensity distributions for three different wavelengths from the red, green and blue parts of the visible spectrum. 
The presented examples demonstrated the possibility of generating color images corresponding to the sum of the generated monochromatic intensity distributions at the operating wavelengths. 
In particular, the presented numerical simulation results shown that using a cascaded DOE consisting of three elements, it is possible to generate a complex color image of parrots with an energy efficiency exceeding 95\% and a root-mean-square error as low as 4.2\%.

\section{Acknowledgments}

This work was funded by the Russian Science Foundation (grant 24-19-00080; development of the gradient method, design and investigation of the cascaded spectral DOEs) and carried out within the state assignment of NRC “Kurchatov Institute” (development of the software implementing the angular spectrum method for simulating cascaded spectral DOEs).

\bibliographystyle{elsarticle-num} 
\bibliography{Parrot}

\end{document}